\documentclass[review]{elsarticle}

\usepackage{lineno,hyperref}
\modulolinenumbers[5]

\journal{Journal of \LaTeX\ Templates}

\begin{document}

\begin{frontmatter}

\title{Quartetting in odd-odd self-conjugate nuclei}

\author[address1]{M. Sambataro\corref{correspondingauthor}}
\cortext[correspondingauthor]{Corresponding author}
\ead{michelangelo.sambataro@ct.infn.it}

\author[address2]{N. Sandulescu}
\ead{sandulescu@theory.nipne.ro}

\address[address1]{Istituto Nazionale di Fisica Nucleare - Sezione di Catania, Via S. Sofia 64, I-95123 Catania, Italy}
\address[address2]{National Institute of Physics and Nuclear Engineering, P.O. Box MG-6, Magurele, Bucharest, Romania}

\begin{abstract}
We provide a description of odd-odd self-conjugate nuclei in the $sd$ shell in a formalism of collective quartets and pairs.  Quartets are four-body structures carrying isospin $T=0$ while pairs can have either $T=0$ or $T=1$. Both quartets and pairs are labeled by the angular momentum $J$ and they are chosen so as to describe the lowest states of $^{20}$Ne (quartets) and the lowest $T=0$ and $T=1$ states of $^{18}$F (pairs).  We carry out configuration interaction calculations in spaces built by one quartet and one pair for $^{22}$Na and by two quartets and one pair for $^{26}$Al. The spectra that are generated are in good agreement with the shell model and experimental ones. These calculations confirm the relevance of quartetting in the structure of $N=Z$ nuclei that had already emerged in previous studies of the even-even systems and highlight the role of $J>0$ quartets in the composition of the odd-odd spectra.
\end{abstract}


\end{frontmatter}

\nolinenumbers

\section{Introduction}
A distinctive feature of self-conjugate nuclei is that of carrying an equal number of protons and neutrons distributed over the same single particle levels. In these nuclei, owing to the charge independence of the nuclear interaction, the isovector proton-neutron ($pn$) pairing is expected to come into play on equal footing as the like-particle proton-proton and neutron-neutron pairing of the more common $N>Z$ nuclei. In addition, $pn$ pairing is also expected to occur in an isoscalar form. The competition between these two types of $pn$ pairing in $N=Z$ nuclei has been matter of great debate in recent years (for a recent analysis on the subject see Ref. \cite{fraue}).

In the above context, particular attention has been addressed to odd-odd $N=Z$ nuclei \cite{vogel,augusto,sagawa,kanada}. These nuclei exhibit a peculiar coexistence of isospin $T=0$ and $T=1$ states at very low excitation energies. This feature is clearly visible in Fig. 1 which shows the experimental positive-parity spectra of the lightest odd-odd $N=Z$ nuclei in the $sd$ shell, namely $^{18}$F, $^{22}$Na and $^{26}$Al. By focusing on the low-lying states within the circles, one may notice that: a), there is always a $T=1,J=0$ state which coexists with some $T=0$ states; b), the energy of this $T=1,J=0$ state decreases with increasing the mass of the nucleus up to becoming almost degenerate with the $T=0$ ground state in $^{26}$Al; c), the low-lying $T=0$ states form a group of three states with angular momenta $J=1,3,5$ (with the only exception of $^{22}$Na which exhibits an ``intruder'' $J=4$ state) which one after the other become the ground state of the system. The nuclear shell model represents a powerful tool to study $sd$ shell nuclei (see, for instance, the exhaustive comparison of experimental spectra and theoretical spectra obtained with the USD/A/B interactions in Ref. \cite{brown}) and, as it will be seen more in detail in the following, accounts well for all the above features. However, the complexity of the shell model wave function is such not to allow a simple description of these features. Providing such a simple description represents the basic goal of the present paper.

We have recently carried out an analysis of the $pn$ pairing in even-even $N=Z$ nuclei
both in the isovector and in the isoscalar channels \cite{qcm1,qcm2,qcm3,sasat1,sasaj0,sasat01}. This analysis has evidenced, on the one hand, that a description of the ground state correlations induced by this interaction in terms of a condensate of collective pairs (of various form \cite{sasat01}) is not satisfactory and, on the other hand, that these correlations can be accounted for to a high degree of precision by approximating the ground state as a product of identical $T=0$ quartets. $T=0$ quartets are four-body correlated structures formed by two protons and two neutrons and, in the case of a spherical mean field, they are also characterized by a total angular momentum $J=0$. One needs to remark that quartets have a long history in nuclear structure 
\cite{flowers,arima,yamamura,catara,hasegawa,hasegawa2,dobes,zelevinsky,fu} but their
complexity has undoubtedly represented a hindrance to the development of quartet models. We have also explored a more sophisticated approximation which consists in letting the quartets to be all distinct and we have verified that it leads to basically exact results in the case of the $pn$ isovector pairing in deformed systems \cite{sasat1}. In all cases the quartets have been constructed variationally for each nucleus.

The $pn$ pairing is a key ingredient of the nuclear force for $N=Z$ nuclei but it is nonetheless only a part of it. In the presence of a full Hamiltonian, other quartets are reasonably expected to come into play besides the $T=0,J=0$ ones emerging from the analysis of the $pn$ pairing. This has been verified in a recent analysis of even-even nuclei in the $sd$ shell which has evidenced a significant role of $T=0,J=2,4$ quartets in the low-lying states of $^{24}$Mg and $^{28}$Si \cite{prl}. The quartets employed in this analysis have not been constructed variationally, as in the quoted works on $pn$ pairing \cite{qcm1,qcm2,qcm3,sasat1,sasaj0,sasat01}, owing to the difficulty in applying this procedure in the presence of quartets of various nature. $T=0$ quartets have been instead simply assumed to represent the lowest states of $^{20}$Ne (two protons and two neutrons outside the $^{16}$O core). Once fixed, these quartets have been no longer modified throughout the calculations. A similar criterion has been adopted also for the $T=1$ and $T=2$ quartets which have been employed in an analysis of the whole isobaric chain of $A=24$ nuclei \cite{prl}. These quartets have been associated with the lowest levels of $^{20}$F $(T=1)$ and $^{20}$O $(T=2)$. No need for $T\neq 0$ quartets in the structure of the even-even $N=Z$ nuclei $^{24}$Mg and $^{28}$Si has been found due to the large gap in energy existing between these and the $T=0$ quartets.

The analysis of odd-odd $N=Z$ nuclei in the $sd$ shell that we are going to illustrate is fully inspired to our previous work on even-even $N=Z$ nuclei in the same shell. We assume that an odd-odd nucleus can be described by resorting to two distinct families of building blocks, one formed by collective $T=0$ quartets and the other by collective pairs. These latter  can have either $T=0$ or $T=1$ depending on the isospin of the state that we want to represent. More precisely, we assume that any state with isospin $T$ of an odd-odd nucleus can be represented as a superposition of products of one or more $T=0$ quartets and one extra pair with isospin $T$. Therefore, within this scheme, the total isospin of the state coincides with that of the pair. Based on the conclusions of Ref. \cite{prl}, we involve in the calculations only three $T=0$ quartets, namely the $J=0,2,4$ quartets describing the lowest three states of $^{20}$Ne. In analogy, as collective pairs with isospin $T$, we assume those describing the lowest three states with that isospin in $^{18}$F (one proton and one neutron outside the $^{16}$O core). These states are characterized by angular momenta $J=1,3,5$ for $T=0$ (see Fig. 1) and $J=0,2,4$ for $T=1$. Once fixed, these quartets and pairs are no longer modified. 

The manuscript is structured as follows. In Section 2, we describe the formalism. In Section 3, we present the results. Finally, in Section 4, we give the conclusions.

\section{The formalism}
We work in a spherically symmetric mean field and label the single-particle states by
$i\equiv \{n_i,l_i,j_i\}$, where the standard notation for the orbital quantum numbers is used. The  $T=0$ quartet creation operator is defined as
\begin{equation}
Q^+_{JM}=\sum_{i_1j_1J_1}\sum_{i_2j_2J_2}\sum_{T'}
q_{i_1j_1J_1,i_2j_2J_2,{T'}}
[[a^+_{i_1}a^+_{j_1}]^{J_1{T'}}[a^+_{i_2}a^+_{j_2}]^{J_2{T'}}]^{JT=0}_{M},
\end{equation}
where $a^+_i$ creates a fermion in the single particle state $i$.
No restrictions on the intermediate couplings $J_1T'$ and $J_2T'$ are introduced. Similarly, the pair creation operator is defined as 
\begin{equation}
P^+_{JM,TT_z}=\sum_{ij}p_{ij}
[a^+_{i}a^+_{j}]^{JT}_{MT_z}.
\end{equation}
In the above expressions $M(T_z)$ stands for the projection of $J(T)$.
The coefficients $q_{i_1j_1J_1,i_2j_2J_2,{T'}}$ and $p_{ij}$ are fixed by carrying out shell model calculations for $^{20}$Ne (quartets) and for $^{18}$F (pairs). The interaction that is used throughout this paper is the USDB interaction \cite{usdb}. Once the quartets and the pairs have been fixed, we perform configuration interaction calculations in the space (we adopt the $m$-scheme)
\begin{equation}
\{Q^+_{J_1M_1}P^+_{JM,T0}|0\rangle \}
\end{equation}
for $^{22}$Na and in the space 
\begin{equation}
\{Q^+_{J_1M_1}Q^+_{J_2M_2}P^+_{JM,T0}|0\rangle \}
\end{equation}
for $^{26}$Al, with $|0\rangle$ representing the reference vacuum. 

\section{Results}
Fig. 2 illustrates how the spectrum of the lowest states of $^{22}$Na (just those within the circle of Fig. 1) evolves by progressively including in the configuration space (3) the three $T=0$ quartets while keeping fixed the presence of all $T=0$ or $T=1$ pairs. A number of things are worthy noticing: a), coupling the $T=0,J=1,3,5$ pairs simply to a $T=0,J=0$ quartet (the case Q(J=0) in Fig. 2) generates the wrong order of $T=0$ states (as compared to the experimental spectrum of Fig. 1 and to the shell model result that will be discussed below); b), adding the $J=2$ quartet (the case Q(J=0,2)) results in the appearance of the ``intruder'' $J=4$ state mentioned above but leaves a wrong ordering of the $T=0$ states; c), further including the $J=4$ quartet (case Q(J=0,2,4)) the spectrum acquires the right structure. In passing from Q(J=0) to Q(J=0,2,4) one can also observe a significant increase of the ground state correlation energy. This quantity is defined as the difference between the ground state energies that are calculated with and without interaction and it is indicated by the number below each ground state. The number in parenthesis gives the relative error of the correlation energy with respect to the corresponding shell model value.

In Fig. 3 we compare the positive-parity experimental spectrum of $^{22}$Na up to an energy of about 4 MeV (EXP) with the spectrum obtained within the present quartet formalism (QM) and with the shell model result (SM). The agreement between QM and SM spectra looks satisfactory over the whole range of energy for both $T=0$ and $T=1$ states. We also observe that the calculated $T=1$ spectrum, characterized by an isospin projection $T_z=0$, is by construction identical to what would be found for $T_z=1$ owing to the isospin symmetry of the interaction and it can therefore be directly compared with the experimental spectrum of $^{22}$Ne $(Z=10,N=12)$. This comparison is shown in Fig. 4 up to an energy of about 6 MeV. A good agreement is found also in this case. 

In Fig. 2 we have investigated how the presence of the various $T=0$ quartets affects the spectrum of the very low states in $^{22}$Na. Before moving to an analysis of $^{26}$Al, we shortly outline the results of a similar investigation concerning the pairs. By keeping intact the family of $T=0,J=0,2,4$ quartets and subtracting every time one of the $T=0, J=1,3,5$ pairs, in no case one generates a properly ordered $T=0$ spectrum of $^{22}$Na. Letting out either the $J=3$ or the $J=5$ pair results in a $J=1$ ground state while excluding the $J=1$ pair generates a correct $J=3$ ground state but shifts the lowest $J=1$ state to a high energy (1.75 MeV).  The set of $J=1,3,5$ pairs therefore represents the minimal set of $T=0$ pairs able, in our approach, to guarantee the correct form of the low-lying $T=0$ spectrum of $^{22}$Na. Things go somewhat differently for the $T=1$ spectrum. In this case, still keeping intact the family of $T=0,J=0,2,4$ quartets and subtracting every time one of the $T=1,J=0,2,4$ pairs, one always observes a low-lying $T=1$ spectrum consisting of three states with, in ascending order, $J=0,2,4$. Significant variations can be observed, however, in this case in the relative position of these $T=1$ states with respect to the $T=0$ ones as well as in the spectrum at higher energies. The choice of the full set of $T=1,J=0,2,4$ pairs emerges from this analysis as the most appropriate one.

Making use of the same sets of quartets and pairs employed for $^{22}$Na we have carried out configuration interaction calculations for $^{26}$Al. The comparison between theoretical and experimental spectra is shown in Fig. 5.
The agreement is satisfactory also in this case although one may notice that three $T=0$ states (with $J=1,2,3$) around 2 MeV  are missing in the QM spectrum in comparison with the SM one. This suggests that the inclusion of some extra element either in the set of quartets or in that of $T=0$ pairs would be welcome. Similarly to what already observed in the case of $^{22}$Na, the $T=1$ spectrum can be directly compared with that of $^{26}$Mg $(Z=12,N=14$). This comparison is shown in Fig. 6. Some deviations with respect to the SM spectrum can be observed around 5 MeV which, these too, point to the need of enlarging the configuration space for a more accurate description of the spectrum in this range of energy.

\section{Conclusions}
In this paper we have provided a description of the odd-odd $N=Z$ nuclei
$^{22}$Na and $^{26}$Al in subspaces of the shell model space built in terms of two families of building blocks: $T=0$ collective quartets and $T=0$ or $T=1$ collective pairs.
These quartets and pairs have been assumed to represent the lowest three states of $^{20}$Ne (quartets) and the lowest three $T=0$ and $T=1$ states of $^{18}$F (pairs). We have verified that this choice of quartets and pairs guarantees a good description of the low-lying $T=0$ and $T=1$ states in $^{22}$Na and $^{26}$Al. It is worthy remarking that, with this choice, the size of the configuration spaces (3) and (4) is by far smaller than that of the corresponding shell model spaces (for $^{26}$Al, for instance, the space (4) counts at most 238 states while 63,094 is the corresponding dimension of the shell model space in the $m$-scheme). Reproducing the basic features of the complex spectra of these odd-odd nuclei in such reduced spaces provides a strong support to the validity of the present approximation scheme. Besides allowing a comprehension of these nuclei that is much simpler and more intuitive than that provided by the shell model, these calculations confirm the central role of  quartetting in $N=Z$ nuclei that had already been evidenced in our previous studies of the even-even systems \cite{qcm1,qcm2,qcm3,sasat1,sasaj0,sasat01,prl}. With respect to these previous works, it appears  even more clearly the role of $T=0,J>0$ quartets. Their presence has not only significant effects on the ground state correlation energy (as already remarked in Ref. \cite{prl}) but it also turns out crucial to generate the proper spectrum of the odd-odd nucleus. An even more direct confirmation of the role played by these quartets could arise from a study of odd $N=Z+1$ systems that we leave for future work. 

{\it Acknowledgments} 
We thank D. Gambacurta for useful discussions. 
This work was supported by the Romanian National Authority for Scientific Research,
CNCS UEFISCDI,  Project Number PN-II-ID-PCE-2011-3-0596. 

\section*{References}

\bibliography{bibliog}

\begin{thebibliography}{10}
\expandafter\ifx\csname url\endcsname\relax
  \def\url#1{\texttt{#1}}\fi
\expandafter\ifx\csname urlprefix\endcsname\relax\def\urlprefix{URL }\fi
\expandafter\ifx\csname href\endcsname\relax
  \def\href#1#2{#2} \def\path#1{#1}\fi

\bibitem{fraue}
S.~Frauendorf, A.~Macchiavelli, Progr. Part. Nucl. Phys. 78 (2014) 24.

\bibitem{vogel}
P.~Vogel, Nucl. Phys. A 662 (2000) 148.

\bibitem{augusto}
A.~O. Macchiavelli, P.~Fallon, R.~M. Clark, M.~Cromaz, M.~A. D. R.~M. Diamond,
  G.~J. Lane, I.~Y. Lee, F.~S. Stephens, C.~E. Svensson, K.~Vetter, D.~Ward,
  Phys. Rev. C 61 (1979) 041303(R).

\bibitem{sagawa}
H.~Sagawa, C.~L. Bai, G.~Col{\` o}, Phys. Scr. 91 (2016) 083011.

\bibitem{kanada}
H.~Morita, Y.~Kanada-En'yo, arXiv:1604.07131.

\bibitem{brown}
http://www.nscl.msu.edu/~brown/resources/resources.html.

\bibitem{qcm1}
N.~Sandulescu, D.~Negrea, J.~Dukelsky, C.~W. Johnson, Phys. Rev. C 85 (2012)
  061303(R).

\bibitem{qcm2}
N.~Sandulescu, D.~Negrea, C.~W. Johnson, Phys. Rev. C 86 (2012) 041302(R.

\bibitem{qcm3}
D.~Negrea, N.~Sandulescu, Phys. Rev. C 90 (2014) 024322.

\bibitem{sasat1}
M.~Sambataro, N.~Sandulescu, Phys. Rev. C 88 (2013) 061303(R).

\bibitem{sasaj0}
M.~Sambataro, N.~Sandulescu, C.~Johnson, Phys. Lett. B 740 (2015) 137.

\bibitem{sasat01}
M.~Sambataro, N.~Sandulescu, Phys. Rev. C 93 (2016) 054320.

\bibitem{flowers}
B.~H. Flowers, M.~Vujicic, Nucl. Phys. A 49 (1963) 586.

\bibitem{arima}
A.~Arima, V.~Gillet, Ann. of Phys. 66 (1971) 117.

\bibitem{yamamura}
J.~Eichler, M.~Yamamura, Nucl. Phys. A 182 (1972) 33.

\bibitem{catara}
F.~Catara, J.~M. Gomez, Nucl. Phys. A 215 (1973) 85.

\bibitem{hasegawa}
M.~Hasegawa, S.~Tazaki, R.~Okamoto, Nucl. Phys. A 592 (1995) 45.

\bibitem{hasegawa2}
M.~Hasegawa, S.~Tazaki, Nucl. Phys. A 633 (1998) 266.

\bibitem{dobes}
J.~Dobes, S.~Pittel, Phys. Rev. C 57 (1998) 688.

\bibitem{zelevinsky}
R.~A. Sen'kov, V.~Zelevinsky, Phys. Atom. Nucl. 74 (2011) 1267.

\bibitem{fu}
G.~J. Fu, Y.~Lei, Y.~M. Zhao, S.~Pittel, A.~Arima, Phys. Rev. C 87 (2013)
  044310.

\bibitem{prl}
M.~Sambataro, N.~Sandulescu, Phys. Rev. Lett. 115 (2015) 112501.

\bibitem{usdb}
B.~A. Brown, W.~A. Richter, Phys. Rev. C 74 (2006) 034315.

\bibitem{nndc}
http://www.nndc.bnl.gov.

\end{thebibliography}

\newpage
\begin{figure}[ht]
\begin{center}
\includegraphics[width=4.2in,height=3.0in,angle=0]{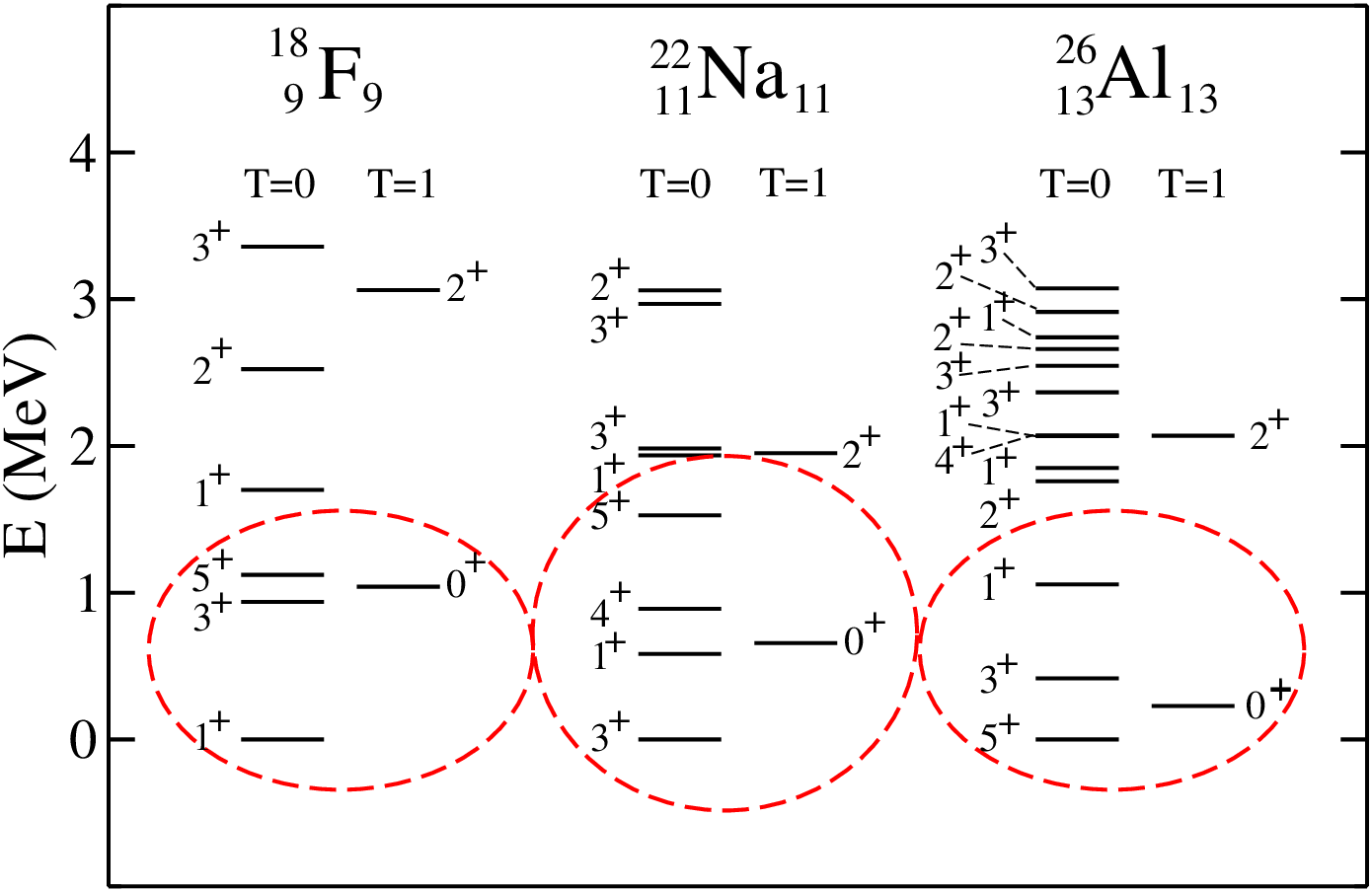}
\caption{
Experimental positive-parity spectra of $^{18}$F, $^{22}$Na and $^{26}$Al (data from National Nuclear Data Center \cite{nndc}).}
\end{center}
\label{figure1}
\end{figure}

\newpage
\begin{figure}[ht]
\begin{center}
\includegraphics[width=4.2in,height=3.0in,angle=0]{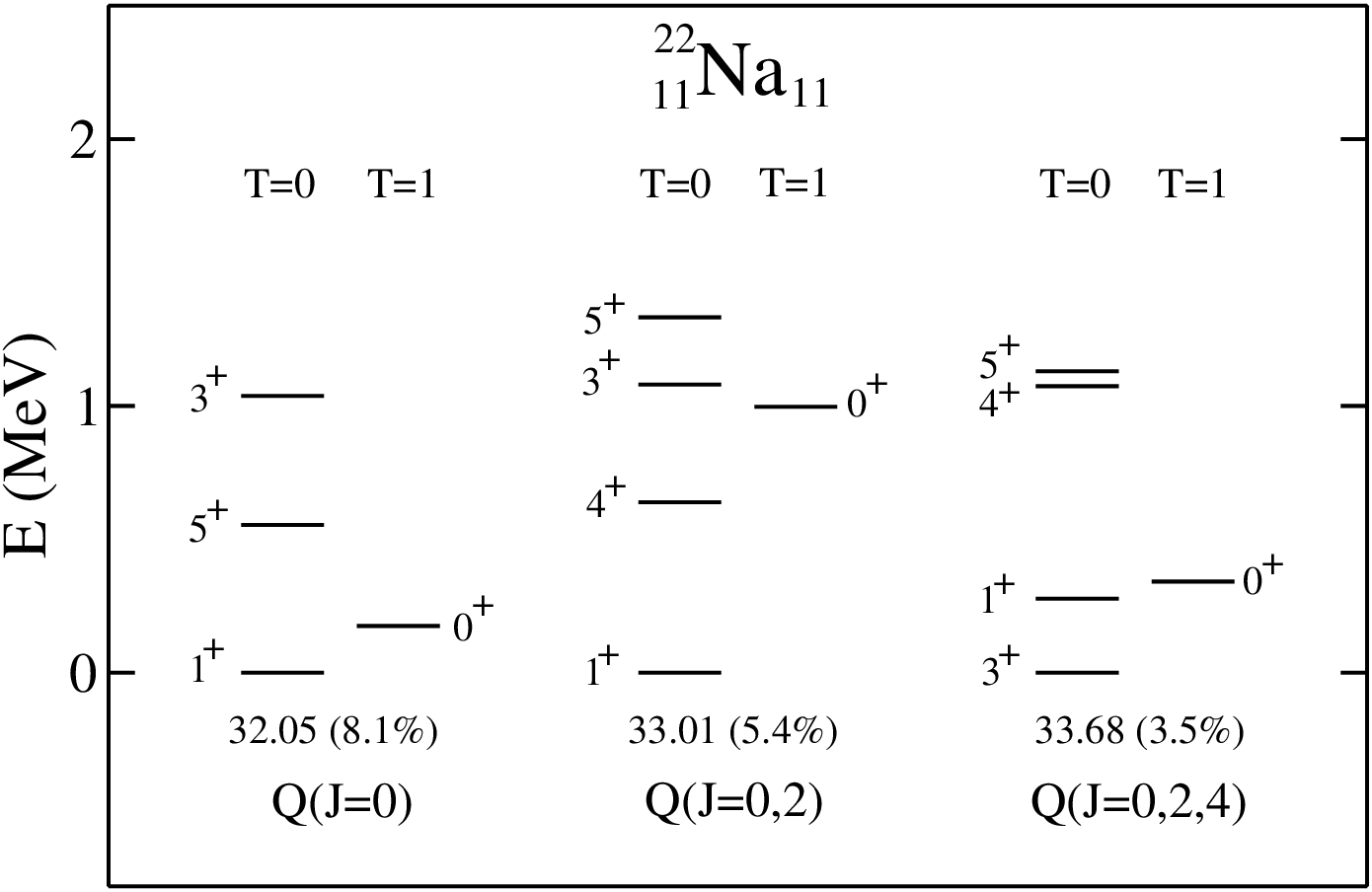}
\caption{
Low-lying states of $^{22}$Na in correspondence with different configuration spaces (3).
Q(J=0) refers to the use of only a $T=0,J=0$ quartet, Q(J=0,2) to the inclusion of $T=0,J=0,2$ quartets and Q(J=0,2,4) to the full set of $T=0,J=0,2,4$ quartets. In all these cases the full sets of pairs described in text are used. The numbers below each spectrum are the ground state correlation energy and (in parenthesis) its relative error with respect to the shell model result.}
\end{center}
\label{figure2}
\end{figure}

\newpage
\begin{figure}[ht]
\begin{center}
\includegraphics[width=4.2in,height=3.0in,angle=0]{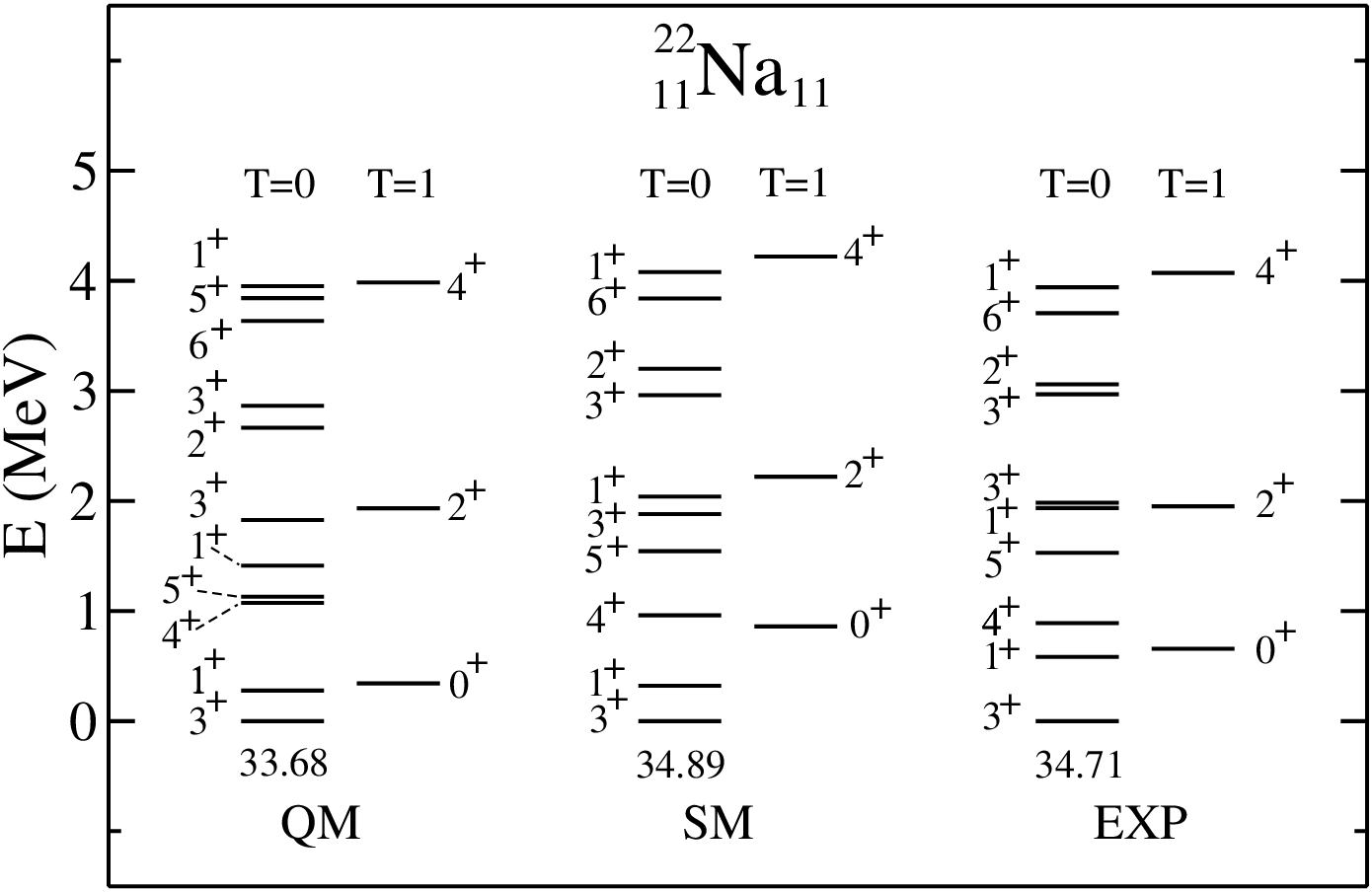}
\caption{
The spectrum of $^{22}$Na obtained with the quartet formalism of this paper (QM) compared with shell model (SM) \cite{brown} and experimental (EXP) \cite{nndc} spectra.
The number below each spectrum is the ground state correlation energy (see text).
}
\end{center}
\label{figure3}
\end{figure}

\newpage
\begin{figure}[ht]
\begin{center}
\includegraphics[width=4.2in,height=3.0in,angle=0]{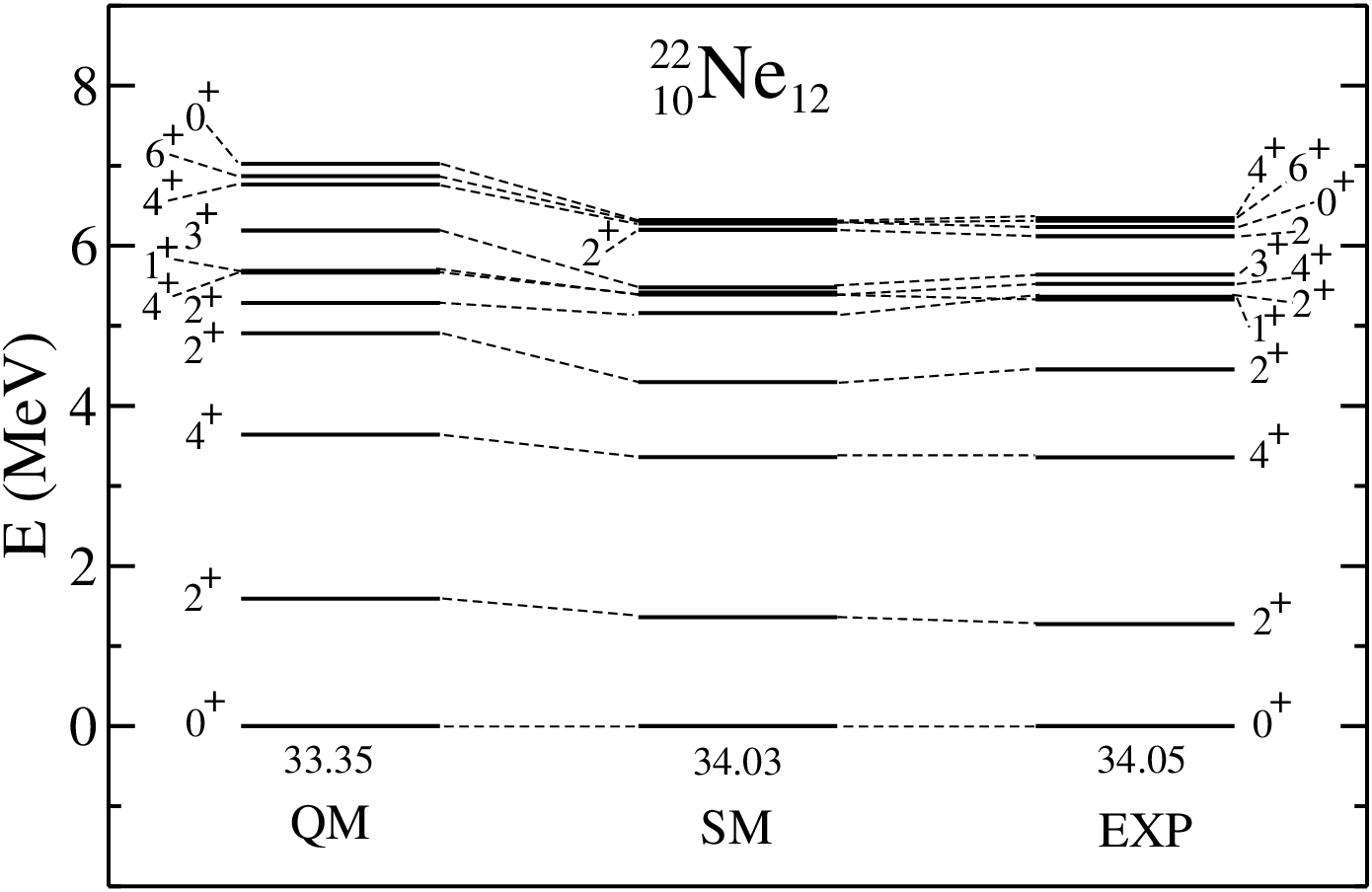}
\caption{
The same as Fig.3 for $^{22}$Ne 
}
\end{center}
\label{figure4}
\end{figure}

\newpage
\begin{figure}[ht]
\begin{center}
\includegraphics[width=4.2in,height=3.0in,angle=0]{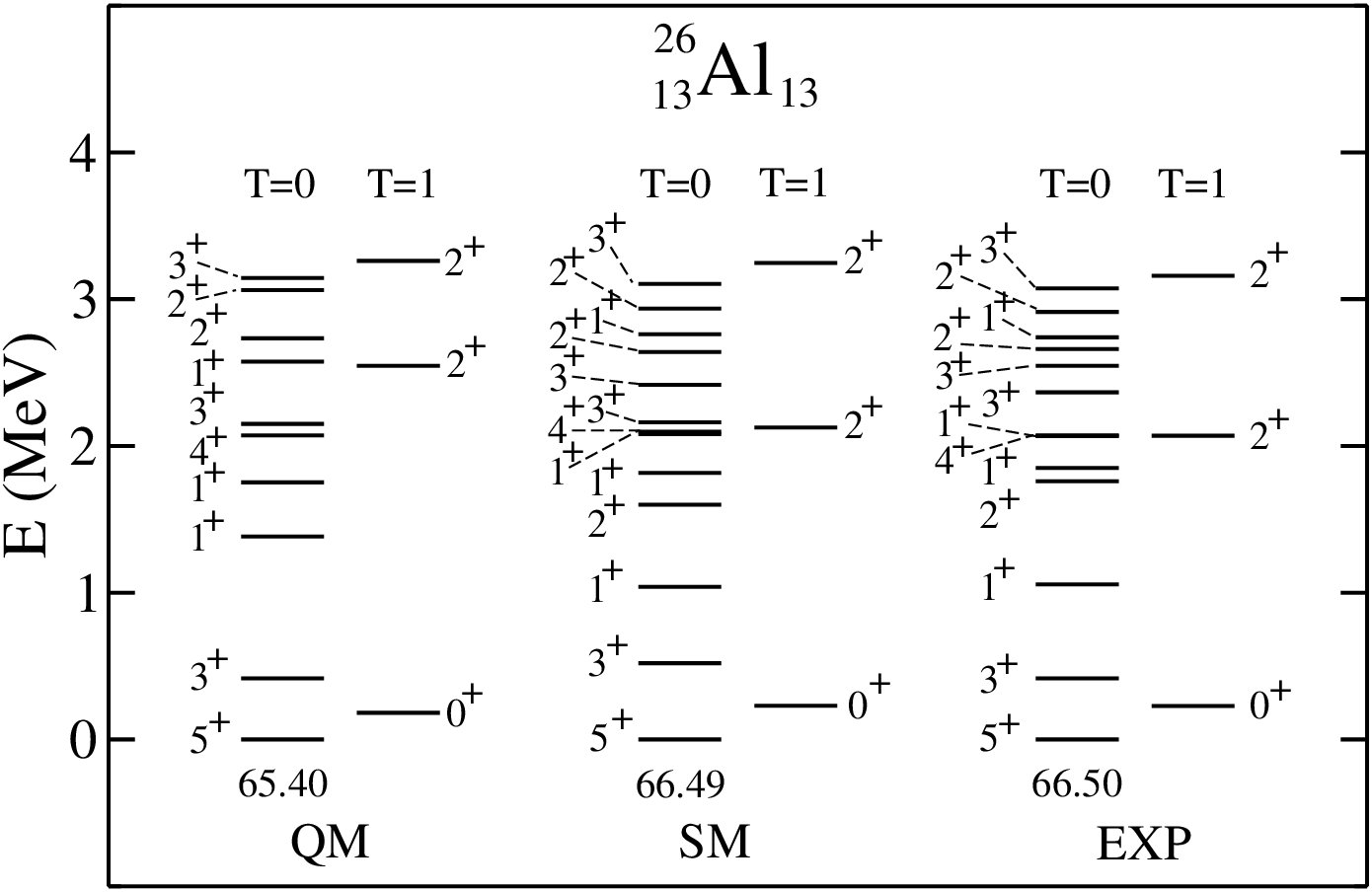}
\caption{
The same as Fig. 3 for $^{26}$Al.
}
\end{center}
\label{figure5}
\end{figure}

\newpage
\begin{figure}[ht]
\begin{center}
\includegraphics[width=4.2in,height=3.0in,angle=0]{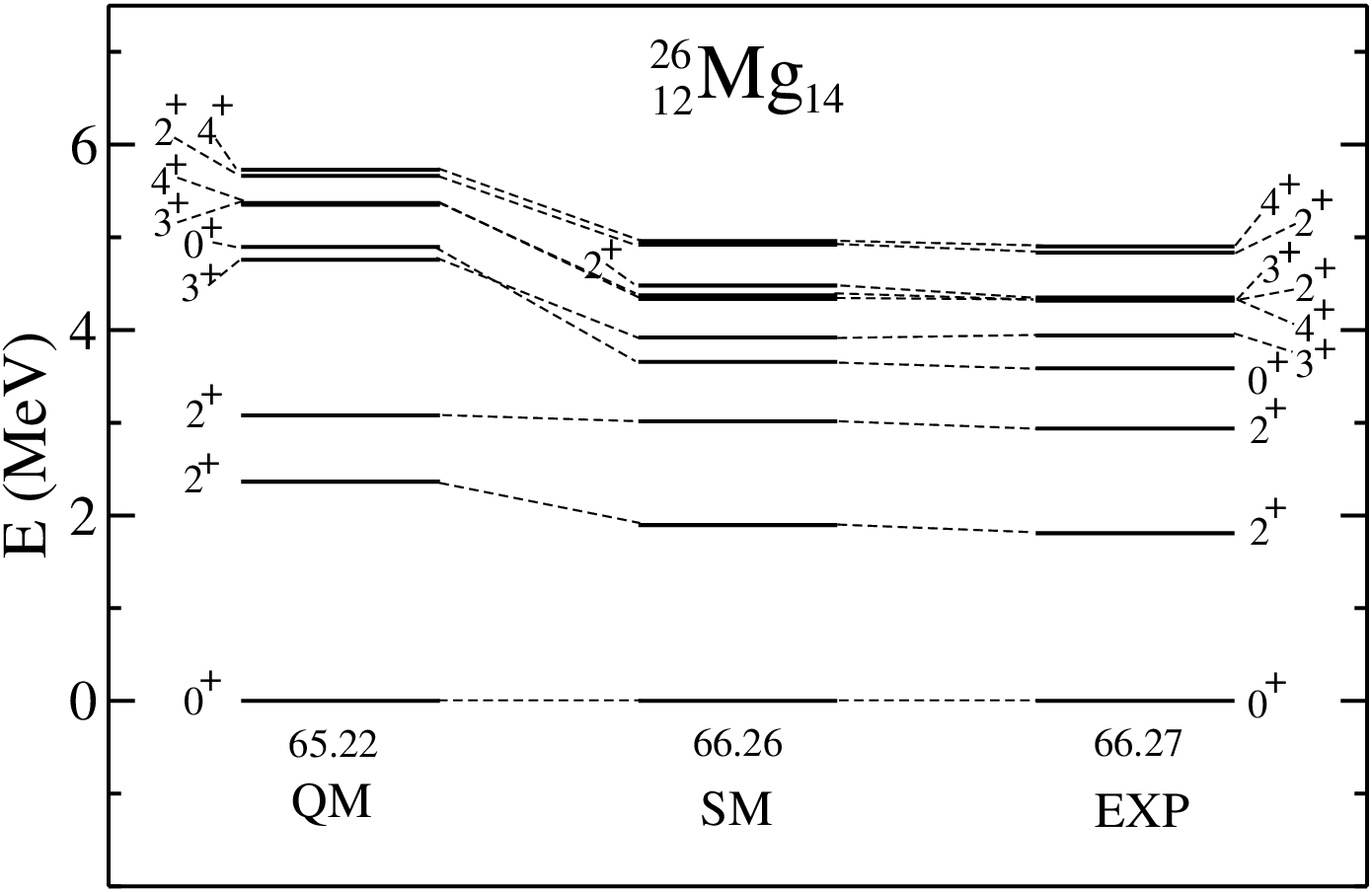}
\caption{
The same as Fig. 3 for $^{26}$Mg.
}
\end{center}
\label{figure6}
\end{figure}

\end{document}